\documentclass[prd, reprint, amsmath, amssymb, aps, superscriptaddress]{revtex4-1}
\pdfoutput=1

\usepackage{appendix}
\usepackage{color}
\usepackage{graphicx}
\usepackage{siunitx}
\usepackage[citecolor=blue, linkcolor=blue, urlcolor=blue, colorlinks=true]{hyperref}

\newcommand{\unit}[1]{\ensuremath{\textrm{#1}}}
\newcommand{\GeV}{\unit{GeV}}
\newcommand{\fb}{\unit{fb}}

\newcommand{\ifb}{\unit{fb}^{-1}}
\newcommand{\iab}{\unit{ab}^{-1}}
\newcommand{\mum}{\SI{}{\micro\metre}}
\newcommand{\tn}[1]{\ensuremath{\textnormal{#1}}}
\newcommand{\KS}{\ensuremath{K_S^0}}
\newcommand{\tagger}{$s$-tagger}
\newcommand{\eg}{\emph{e.g.}}

\begin{document} 

\title{Probing the strange Higgs coupling at \texorpdfstring{$e^+e^-$}{e+e-} colliders using
  light-jet flavor tagging}

\author{J.~Duarte-Campderros}
\email{jorge.duarte.campderros@cern.ch}
\affiliation{Universidad de Cantabria Instituto de Fisica de Cantabria (IFCA) Avda. Los Castros
  s/n, E-39005 Santander, Spain}
\affiliation{European Organization for Nuclear Research (CERN) EP-CMX Dept.  CH-1211 Geneva 23,
  Switzerland}
\author{G.~Perez}
\email{gilad.perez@weizmann.ac.il}
\affiliation{Department of Particle Physics and Astrophysics, Weizmann Institute of Science,
  Rehovot 7610001, Israel}
\author{M.~Schlaffer}
\email{matthias.schlaffer@weizmann.ac.il}
\affiliation{Department of Particle Physics and Astrophysics, Weizmann Institute of Science,
  Rehovot 7610001, Israel}
\author{A.~Soffer}
\email{asoffer@post.tau.ac.il}
\affiliation{School of Physics and Astronomy, Tel-Aviv University, Tel-Aviv 69978, Israel}

\begin{abstract}
  We propose a method to probe the coupling of the Higgs to strange quarks by tagging strange jets
  at future lepton colliders. For this purpose we describe a jet-flavor observable, $J_F$, that is
  correlated with the flavor of the quark associated with the hard part of the jet. Using this
  variable, we set up a strangeness tagger aimed at studying the decay $h\to s\bar{s}$. We determine
  the sensitivity of our method to the strange Yukawa coupling, and find it to be of the order of
  the standard-model expectation.
\end{abstract}

\maketitle

The hierarchical structure of the masses and mixing angles of the fundamental, point-like matter
fields is one of the unsolved mysteries of the standard model (SM) of elementary particles. This
\emph{flavor puzzle} extends beyond the SM, since new-physics models are severely constrained by
measurements of flavor observables. Within the SM, the flavor structure is trivially encoded by the
structure of the Yukawa interactions: the Yukawa coupling of fermion $f$ to the physical Higgs is
given by the ratio of the fermion mass to the Higgs vacuum expectation value, $y_f=m_f/v$. Beyond
the SM, various alternatives motivated by the flavor puzzle predict significant deviations from this
structure (see, \eg, Refs.~\cite{Giudice:2008uua, Delaunay:2013pja, Kagan:2014ila, Dery:2014kxa,
  Bauer:2015fxa, Ghosh:2015gpa}). Now that we have entered the era of
precision Higgs measurements, one can begin to tackle this question experimentally, opening a new
direction in flavor physics.

Experimental tests of the Yukawa interactions are performed in terms of the signal strength $\mu_P$,
defined as the ratio of the measured production cross section or branching ratio of the process $P$
to the SM expectation for this quantity. The latest results from the ATLAS~\cite{Aaboud:2018zhk,
  Aaboud:2018urx, Aaboud:2018pen} collaboration are
\begin{equation}
    \label{eq:mu-measured}
    \mu_{tth}=1.32^{+0.28}_{-0.26},~\mu_{bb} = 1.01^{+0.20}_{-0.20},~\mu_{\tau\tau}=1.09^{+0.35}_{-0.30},
\end{equation}
where $P=tth$ refers to the process $pp\to t\bar{t}h$ and $P=bb$ denotes the decay $h\to b\bar{b}$,
etc.  Similarly, the CMS~\cite{Sirunyan:2018koj} collaboration has measured
\begin{equation}
  \mu_{tth} = 1.18^{+0.30}_{-0.27},~\mu_{bb} = 1.12^{+0.29}_{-0.29},~\mu_{\tau\tau} =1.02^{+0.26}_{-0.24}.
\end{equation}
Within the large uncertainties, these results are in good agreement with the SM expectation for the
third-generation fermions, for which the Yukawa couplings are large and thus easier to
measure. Precision will improve as the experiments collect more data.

However, it is important to note that the flavor puzzle is related to the mass hierarchy among all
three generations of up-type quarks ($u$, $c$, $t$), down-type quarks ($d$, $s$, $b$) and charged
leptons ($e$, $\mu$, $\tau$). Therefore, direct measurements of the smaller couplings of the Higgs
to the first two generations of the different sectors are also necessary. So far, only upper bounds
on the corresponding signal strengths have been obtained~\cite{Aaboud:2017ojs, Khachatryan:2014aep,
  Perez:2015aoa, Aaboud:2017xnb},
\begin{equation}
  \begin{aligned}
    \label{eq:yukawa-limit}
    \mu_{\mu\mu}& \lesssim 2.8,  & \mu_{ee} & \lesssim 3.7\times 10^5,\\
    \mu_{cc} & \lesssim 110, & \mu_{ss} &\lesssim 7.2\times 10^8.
  \end{aligned}
\end{equation}

There is a clear path towards measuring the Yukawa couplings of the muon~\cite{Aaboud:2017ojs},
charm quark~\cite{Bodwin:2013gca}, and possibly even the electron~\cite{Greco:2016izi}. For the
light quarks, the situation is much less clear. It is possible that their Yukawa couplings could be
probed exclusively~\cite{Kagan:2014ila} or indirectly~\cite{Brivio:2015fxa, Bishara:2016jga,
  Soreq:2016rae, Yu:2016rvv, Gao:2016jcm}.  However, these methods are characterized by small rates
and/or are subject to large QCD backgrounds that are difficult to control (see,
\eg,~Refs.~\cite{Perez:2015lra, Kamenik:2016tve}).

In this letter we propose using a strangeness tagger, inspired by $Z\to s\bar{s}$ measurements at
the \mbox{DELPHI}~\cite{Abreu:1999cj} and SLD~\cite{Kalelkar:2000ig} experiments, for probing the
Higgs coupling to the strange quark via the $h\to s\bar{s}$ decay. We first present the basic idea
and describe a new jet-flavor variable, $J_F$, for determining the light flavor of a jet or other
collection of particles. Subsequently, we implement $J_F$ for strange jets and combine it with a
simple method for rejecting background from heavy-flavor jets. This combination of a cut on $J_F$
and the heavy-flavor rejection as well as a choice for the parameter intended to select kaons is
denoted as \tagger{} in the following. Finally, we apply this \tagger{} to a collider scenario with
background from other Higgs decays and non-Higgs events.

Since the branching ratio for $h\to s\bar{s}$ is small, the measurement requires a large number of
Higgs bosons in a very clean environment.  Therefore, we limit the discussion to lepton colliders,
having in mind the International Linear Collider (ILC)~\cite{Barklow:2015tja}, the Future Circular
Collider in electron mode (FCC-ee)~\cite{Benedikt:2015ktd}, and the Circular Electron Positron
Collider (CEPC)~\cite{CEPCStudyGroup:2018rmc}.  These proposed colliders would run at a
center-of-mass energy $\sqrt{s}=250~\GeV$, where the cross section for associated Higgs production,
$e^+e^-\to Zh$, peaks at $\sigma\approx210~\fb$ for unpolarized beams~\cite{Mo:2015mza,
  Thomson:2015jda}. We therefore adopt this scenario for this letter.

Flavor tagging relies on the idea that the flavor of a parton that belongs to the hard part of the
event is correlated with properties of the collection of final-state hadrons originating from that
parton. This collection may be the hadrons in a jet or any other part of the event, \eg,~a
hemisphere defined by a particular axis. We use the term \emph{jet} as a generic umbrella term for
both, with \emph{jet flavor} referring to the flavor of the primary parton that gave rise to the
jet. The process of determining jet flavor is referred to as \emph{flavor tagging}.

While QCD conserves flavor, it does impact the particle composition of a jet, thus affecting jet
flavor tagging.  For example, flavor-blind gluon splitting can result in a quark appearing inside
the jet while its partner antiquark does not, resulting in the creation of flavor that is
differential in phase space~\cite{Banfi:2006hf}.  In the case of heavy quarks, this can be mitigated
perturbatively by modifying the clustering algorithm used to construct the jet. This is done either
by undoing the flavor creation by gluon splitting~\cite{Banfi:2006hf} or by allowing the jet to form
a higher representation (beyond fundamental) of the SM global flavor group~\cite{Ilten:2017rbd}.
Here, we attempt to tackle the same issue for the light-flavor quarks.  The price to pay is that we
must give up some perturbative control and hence calculability.  Specifically, the jet-flavor
variable that we define below is safe against soft but not collinear radiation.

While the processes of showering and hadronization degrade the flavor-tagging capability, some of
the correlation between the primary parton and the final-state hadrons remains. To exploit this, we
define a new jet-flavor variable,
\begin{equation}
  \label{eq:JF-definition}
  J_F = \frac{\sum\limits_{H}\mathbf{p}_{H}\cdot\mathbf{\hat{s}}\, R_{H}}{\sum\limits_{H}\mathbf{p}_{H}\cdot\mathbf{\hat{s}}}.
\end{equation}
Here, the sum is over all hadrons inside the jet, $\mathbf{p}_{H}$ is the momentum vector of the
hadron $H$, $R_H$ is its quantum number or numbers in the flavor representation of interest, and
$\mathbf{\hat{s}}$ is the normalized jet axis. In our case of a Higgs boson produced approximately
at rest and undergoing a $h\to jj$ decay, $\left|\sum\mathbf{p}_H\right|\approx m_h/2$, so that the
denominator is nonzero, and $J_F$ is well defined. Here and in the following we use $h\to jj$ to
denote a Higgs decay into a final state of two gluons or a quark--anti-quark pair.

As our focus is on strangeness tagging, we use only the SU(3) flavor composition for evaluating
$R_H$.  For further simplicity, we consider only pions and kaons, which constitute the majority of
final-state hadrons. Since $m_{u,d}\ll m_s\sim \Lambda_{\rm QCD}$, the SU(3) flavor symmetry is
broken in a rather strong way by the strange-quark mass. Therefore, it is enough to consider only
strangeness, assigning $R_{K^\pm}=\mp1$ and $R_{\pi^{\pm,0}}=0$. The opposite sign of $R_{K^\pm}$
for oppositely charged kaons ensures that kaons originating from $g\to s\bar{s}$ contribute
minimally to $J_F$. The presence of neutral kaons weakens the ability to perform strangeness tagging
for two reasons. First, it is impossible to determine whether a neutral kaon has strangeness $1$ or
$-1$. Second, only the $K_S$ decays in a clearly distinguishable manner, while the $K_L$ gives rise
to calorimeter energy deposits that are not unique. Therefore, a reasonable, minimal-bias approach
is to assign $R_{K_S}=1$ or $-1$ to each $K_S$ in the jet, choosing the set of values that minimizes
the value of $|J_F|$.

We test our \tagger{} on $e^+ e^-\to Zh$ events that are generated at $\sqrt{s}=250~\GeV$ and
showered using \textsc{PYTHIA} version 8.219~\cite{Sjostrand:2006za, Sjostrand:2014zea}. For the
purpose of studying background from hadronic $W$ decays, we additionally generate $e^+e^-\to W^+W^-$
events. As a cross check, we also generate events with \textsc{Herwig} version
7.1.4~\cite{Bahr:2008pv, Bellm:2015jjp}, and find good agreement in the relevant kinematic
distributions. Our study is based on truth-level information, without full detector
simulation.  In order for the estimate to be more realistic, we make several assumptions about the
reconstruction performance, as described below, based on the performance of existing and proposed
detectors.

We assume that in the reconstruction of an $e^+e^-\to Zh$ event, the final-state hadrons originating
from the Higgs decay are correctly associated to the Higgs. Therefore, all particles from the $Z$
decay are ignored at the truth level. This is especially justified, since we consider the invisible
decay mode of the $Z$ boson later on.

As discussed above, the crucial final-state hadrons for our \tagger{} are kaons and charged pions,
which can fake charged kaons. We make the following assumptions about their reconstruction.

\emph{Neutral kaons:} Among the neutral kaons and their decays, only \KS{} mesons decaying into two
charged pions can be efficiently reconstructed. This decay is identified from the two pion tracks
that form a displaced vertex.  We require that the \KS{} decay takes place at a (3-dimensional)
distance of between $5~\unit{mm}$ and $1~\unit{m}$ from the interaction point (IP). The lower cutoff
removes prompt background, and the upper one ensures that the decay occurs sufficiently deep in the
detector for the pion tracks to be well reconstructed. For $K_S$ mesons satisfying these
requirements, we assume a total reconstruction efficiency of $85\%$. This is similar to the
performance of the ATLAS detector \cite{ATL-PHYS-PUB-2017-014}.

\emph{Charged kaons:} We assume that a charged pion or kaon is reconstructed with an average
efficiency of 95\%. In reality, an even better reconstruction efficiency is expected
\cite{Abada:2019zxq}. We further take into account the possibility that the detector has a particle
identification (PID) capability, to discriminate pions from kaons. For concreteness we adopt the PID
capability of the IDEA drift chamber with cluster counting, which can separate pion from kaon tracks
by more than 5 standard deviations in the relevant momentum range \cite{Abada:2019zxq}. As a
parameter describing the identification of kaons candidates, we take the selection efficiency,
$\epsilon_{K^\pm}$, of the PID cut from which we obtain the misidentification probability for pions
given a pion-kaon separation.

The event is divided to two hemispheres along the direction of the sphericity
axis~\cite{Parisi:1978eg, Donoghue:1979vi}, and each final-state particle originating from the Higgs
decay is identified with the hemisphere into which its momentum vector points. For each hemisphere
we calculate $J_F$ according to Eq.~(\ref{eq:JF-definition}) with $\mathbf{\hat{s}}$ set to the
sphericity axis.  A simpler strangeness-tagging approach based on the hardest kaon of a hemisphere
was employed in the \mbox{DELPHI}~\cite{Abreu:1999cj} and SLD~\cite{Kalelkar:2000ig} studies of
$Z\to s\bar{s}$ decays.

Higgs decays to bottom or charm quarks are a significant background in the study of $h\to s\bar{s}$,
warranting special consideration. We take advantage of the fact that bottom and charm hadrons have
sufficiently large lifetimes that their decays occur visibly away from the IP.\@ Consequently, a
charged-kaon track produced in a heavy-flavor hadron decay has a non-vanishing transverse impact
parameter $d_0$ with respect to the IP.\@ By contrast, in a strange jet, the selected kaon
candidates are produced promptly. Taking advantage of this, we further suppress the heavy-quark
background by requiring a small $d_0$ value for all reconstructed kaon candidates with lab-frame
momentum larger than $5~\GeV$. The momentum requirement rejects kaons for which $d_0$ is poorly
measured.

We account for the $d_0$ measurement resolution by smearing the true value of $d_0$ for a given
track with a Gaussian distribution. The Gaussian width is taken to be
$\sigma_{d_0}=\sqrt{\Delta_\tn{IP}^2+\Delta_{d_0}^2}$, where $\Delta_\tn{IP}$ is the uncertainty in
the location of the IP, and $\Delta_{d_0}$ is the uncertainty in the location of the track. We
approximate the collider-dependent~\cite{BARISH:2013doa, Boscolo:2017nir} $\Delta_\tn{IP}$ to be
$5~\mum$ and parameterize $\Delta_{d_0}$ as a function of the kaon track momentum $p$ and polar angle
$\theta$ in the laboratory frame~\cite{Behnke:2013lya},
\begin{equation}
  \label{eq:D0_resolution}
  \Delta_{d_0}=\sqrt{\left(5~\mum\right)^2+\left(\frac{10~\GeV}{p\sin^{3/2}\theta}~\mum\right)^2}. 
\end{equation}

Heavy-flavor background can be further reduced by vetoing a kaon candidate if it is part of a jet
that contains additional large-$d_0$ tracks or a multitrack vertex that is displaced from the IP in
a manner consistent with charm or bottom decay. Algorithms relying on such signatures are regularly
used to select heavy-flavor jets~(see, \eg,~Ref.~\cite{Aaboud:2018xwy}). Implementing these
algorithms as a heavy-flavor veto involves experimental aspects that are beyond the scope of our
analysis. In particular, we note that the large-$J_F$ requirement tends to select bottom and charm
hadrons that undergo few-body decays, necessitating careful study of the veto performance as a
function of $J_F$ rather than relying on studies reported in the literature for generic decays.
Further improvements to background rejection can be achieved by using jet-shape variables. Since we
do not implement these additional algorithms, our results on the ability of the \tagger{} to reject
heavy-flavor jets can be taken as conservative.

\begin{figure}[tb]
  \centering
  \includegraphics[width=\linewidth]{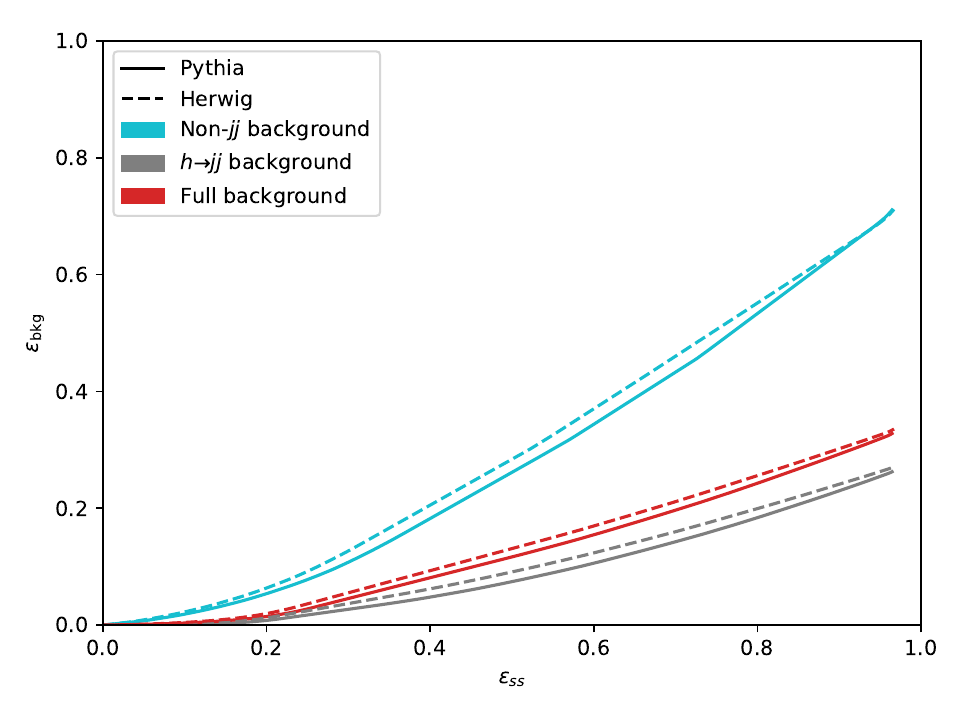}
  \caption{Lowest possible background efficiency for specific background processes vs.~given signal
    efficiency of the \tagger{}.}
  \label{fig:roc}
\end{figure}
For an assessment of the general performance of the \tagger{} we plot in Fig.~\ref{fig:roc} the
lowest achievable \tagger{} efficiency for different background channels vs.\ given efficiencies of
the signal channel.  Here, $h\to jj$ background refers to dijet Higgs decays given SM branching
ratios, Non-$jj$ background is a combination of all other background processes as detailed below,
and Full background refers to the combination of the two background classes with their relative
contribution as given in Ref.~\cite{Bai:2016}.

We study only the signal process $e^+e^-\to Zh$ with $Z\to\nu\bar\nu$, which has been shown to yield
the highest sensitivity to $h\to jj$~\cite{Ono:2013sea, Bai:2016, Abramowicz:2016zbo,
  CEPCStudyGroup:2018ghi}. For an estimation of the background from non-$h\to jj$ events---events
without a Higgs, events where the Higgs does not decay into two gluons or a quark--anti-quark pair,
or events where the Higgs is not produced in association with two neutrinos---we rely on previous
studies~\cite{Ono:2013sea, Bai:2016}. Ref.~\cite{Ono:2013sea} provides a conservative estimate for
the achievable signal to background ratio, as it uses a too small Higgs mass of $120~\GeV$, thus
underestimating the performance of its cut on the invariant dijet mass, $m_{jj}$ of the Higgs
candidate. Moreover, the results were obtained using a cut-and-count technique, whereas a much
better signal-background separation could be obtained by using modern multivariate
methods. Ref.~\cite{Bai:2016} provides an estimate of the performance improvement that can be
expected by using a boosted decision tree (BDT). While Refs.~\cite{Ono:2013sea} and \cite{Bai:2016}
differ significantly in their methods and in the levels of background suppression achieved, what is
important for the current analysis is that they report similar flavor compositions for the
non-$h\to jj$ background. Moreover, Ref.~\cite{Bai:2016} reports separately selection efficiencies
for different $h\to jj$ decay modes, showing that their $h\to jj$ selection efficiency is
independent of the Higgs decay mode.

Refs.~\cite{Ono:2013sea, Bai:2016} used a modular approach, in which they first applied cuts to
separate $h\to {jj}$ from all non-$h\to {jj}$ background events, and then applied a flavor tag on
the selected signal-rich sample to select the desired final-state flavor. We refer to these two sets
of cuts as \emph{preselection} and \emph{flavor cuts}, respectively, and adopt the same modular
approach. In the following we take the backgrounds reported in Ref.~\cite{Ono:2013sea} and
obtain the efficiency of the flavor cuts applied within the \tagger{} as follows.

The main non-$h\to jj$ background is $e^+e^-\to W^+W^-$, where one $W$ decays leptonically, thus
generating missing energy, and the other $W$ decays hadronically. These events account for 67\% of
the non-$h\to jj$ background after the preselection cuts. We obtain the \tagger{} efficiency for
this background from the simulated $e^+e^-\to W^+ W^-$ sample, taking the hadronically decaying $W$
to fake the $h\to jj$ decay.

The remaining non-$h\to jj$ background is from the $e^+e^-\to\nu\bar{\nu}q\bar{q}$ and
$e^+e^-\to q\bar{q}$ processes. In order to obtain the relative flavor composition of the quark
pairs faking the Higgs boson in these processes, we generate them using \textsc{MadGraph5} version
2.6.5~\cite{Alwall:2014hca}. We find that the $u\bar{u}$ and $c\bar{c}$ Higgs candidates from these
backgrounds account for 22\% each, and the $d\bar{d}$, $s\bar{s}$, and $b\bar{b}$ Higgs candidates
make up 19\% each. Since here the faked Higgs has the same partonic final state as the Higgs in the
generated $e^+e^-\to Zh$ samples, we take the \tagger{} efficiency for these background processes
from the $Zh$ samples, accounting for the different relative contributions of the quark flavors.

\begin{figure}[tb]
  \centering
  \includegraphics[width=\linewidth]{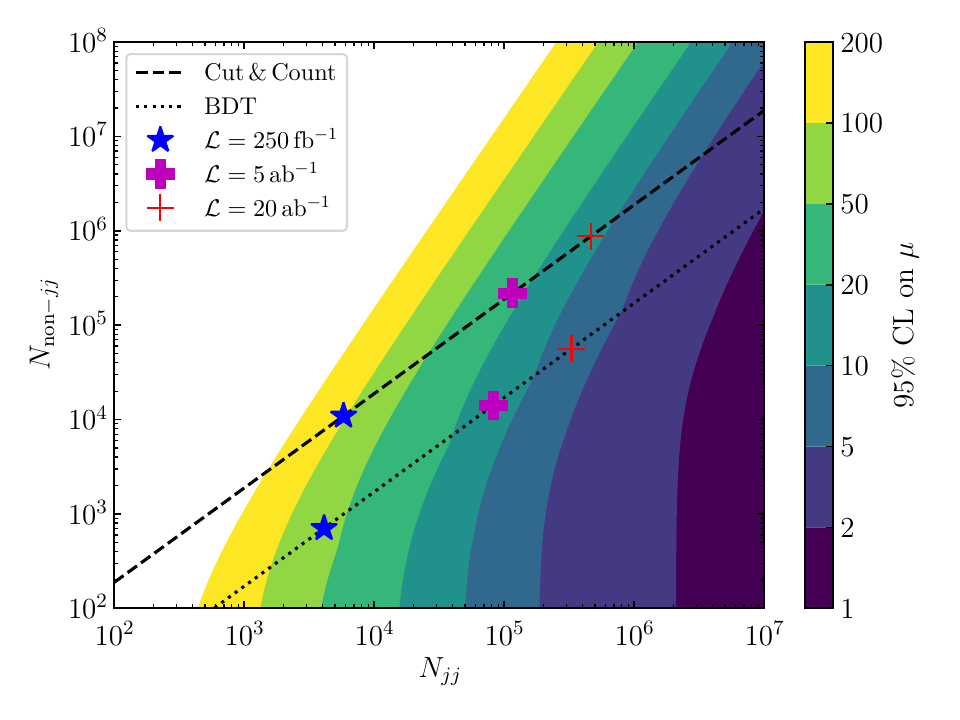}
  \caption{The plot shows on the $x$-axis the number of $h\to jj$ events and on the $y$-axis the
    number of non-$h\to jj$ events, both after the preselection but before the flavor cuts. For
    details see text around Eqs.~\eqref{eq:xaxis} and \eqref{eq:yaxis}. The dashed and dotted lines
    show the constant ratio of $h\to jj$ to non-$h\to jj$ events obtained in a
    cut-and-count~\cite{Ono:2013sea} and BDT analysis~\cite{Bai:2016}, respectively. Integrated
    luminosities for several collider benchmarks. The colored contours show the best limit on the
    signal strength $\mu_{ss}$ obtained after applying the \tagger{} with optimized cuts and
    PID parameter.}
  \label{fig:bestWP}
\end{figure}

Our results are summarized in Fig.~\ref{fig:bestWP} in terms of the number of non-$h\to jj$
events ($N_{{\rm non-}jj}$) vs.~the number of $h\to jj$ events ($N_{jj}$), obtained \emph{after} the
preselection but \emph{before} the flavor cuts. To be precise,
\begin{equation}
  \label{eq:xaxis}
 N_{jj} = \mathcal{L}\,\sigma_h\,\mathcal{B}(h\to jj)\,\epsilon_{{jj}},
\end{equation}
where $\mathcal{L}$ is the integrated luminosity, $\sigma_h$ is the production cross section for
$e^+e^-\to\nu\bar\nu h$ via both $Zh$ and $WW$-fusion, $\mathcal{B}(h\to jj)$ is the total branching
fraction for $h\to jj$ processes, and $\epsilon_{{jj}}$ is the efficiency for such an event to to
pass the preselection criteria. Similarly, the number of non-$h\to jj$ events after preselection is
\begin{equation}
  \label{eq:yaxis}
  N_{\textrm{non-}jj}=\mathcal{L}\sum_{i\in\textrm{non-}jj}\sigma_{i} \epsilon_{i},
\end{equation}
where the sum is over all the non-$h\to jj$ processes, $\sigma_i$ is the cross section for the
process, including relevant branching fractions, and $\epsilon_i$ is the efficiency for events of
the process to pass the preselection criteria.  Fig.~\ref{fig:bestWP} shows diagonal dashed and
dotted lines of constant $N_{jj}/N_{{\rm non-}jj}$, with the values of this ratio being those
obtained using the cut-and-count~\cite{Ono:2013sea} and BDT~\cite{Bai:2016} techniques,
respectively. On these lines, we show the points corresponding to integrated luminosities of $0.25$,
$5$, and $20~\iab$.

For a given value of $N_{jj}$, we determine the number of $h\to jj$ background events for the
dominant Higgs decays $h\to b\bar{b}$, $h\to c\bar{c}$, and $h\to gg$ given the SM Higgs branching
fractions~\cite{deFlorian:2016spz}.  We also determine the number of $h\to s\bar{s}$ signal events,
taking
$\mathcal{B}(h\to s\bar{s}) = \mathcal{B}(h\to c\bar{c})\left(m_s/m_c\right)^2\approx 2.3\times
10^{-4}$, where we evaluate the charm and strange quark masses at the Higgs mass scale using
\textsc{RunDec} version 3.0~\cite{Herren:2017osy}. Given this composition of the $h\to jj$ events
and that of the non-$h\to jj$ events, we iterate over points in the plane of Fig.~\ref{fig:bestWP}
and evaluate the number of signal and background events for varying $\epsilon_{K^\pm}$ and 
cuts on $d_0$, and $J_F$. We then select the cuts that yield the strongest upper limit on the signal
strength $\mu_{ss}$ at 95\% confidence level, and present the upper limits as contours in
Fig.~\ref{fig:bestWP}.

For small values of the ratio $N_{jj}/N_{\textrm{non-}jj}$---around and above the cut-and-count
line---very loose flavor cuts yield the strongest, yet weak limits on $\mu_{ss}$. For the larger
values of $N_{jj}/N_{\textrm{non-}jj}$ obtained by the BDT analysis~\cite{Bai:2016}, $h\to jj$
becomes the dominant background, so that tighter flavor cuts significantly increase the sensitivity
to the $h\to s\bar{s}$ decay when the luminosity is sufficiently large. For the $5~\iab$ benchmark on
the BDT line the best results are still obtained for a working point with loose cuts yielding a
signal efficiency of $\epsilon_{ss}\approx96\%$ and correspondingly background efficiencies of
$\epsilon_\textrm{bkg}^{h\to jj}\approx26\%$ and
$\epsilon_\textrm{bkg}^{\textrm{non-}jj}\approx70\%$. For the $20~\iab$ benchmark, a tight flavor tag
is preferred for the \textsc{PYTHIA} set, yielding $\epsilon_{ss}\approx 14\%$,
$\epsilon_\textrm{bkg}^{h\to jj}<1\%$, and $\epsilon_\textrm{bkg}^{\textrm{non-}jj}\approx3\%$. With
these points we obtain an upper limit on the signal strength of $\mu_{ss}\lesssim 14$ and
$\lesssim 7$ for integrated luminosities of $5$ and $20~\iab$, respectively.  The limit is
weakened to $\mu_{ss}\lesssim 60$ for an integrated luminosity of $250~\ifb$.

The limits in Fig.~\ref{fig:bestWP} do not include systematic uncertainties. However, given the
small number of signal events, the experimental analysis will be statistics limited. For an estimate
of the systematic uncertainty on our predictions we compare the values of $\mu_{ss}$ obtained with
the data sets generated with \textsc{PYTHIA} and \textsc{Herwig}, respectively, and find them to
differ by less than 2\% for the three luminosity benchmarks indicated on the BDT and cut-and-count
lines. Only when the non-$h\to jj$ background is reduced to less than about 10\% of the total
background does the difference increase to 15-20\%. By studying the \tagger{} performance with the
large sample collected by a future collider, particularly in $Z$ decays and 3-jet events, this
uncertainty will be greatly reduced.

We have shown that even with a simple tagging algorithm a very large improvement could be obtained
at a high-luminosity lepton collider, making it sensitive to the strange Yukawa coupling at the
level of several times the SM prediction.  This is to be compared to the estimated limit obtainable
at the HL-LHC ranging from $\mu_{ss}^\textnormal{HL-LHC} \lesssim 200$ from a global fit to
$\mu_{ss}^\textnormal{HL-LHC}\lesssim 10^7$ from the exclusive decay \cite{Cepeda:2019klc}.

\begin{acknowledgments}
  We thank Marumi Kado and Fabio Maltoni for useful discussions.  The
  work of GP is supported by grants from the BSF, ERC, ISF, the
  Minerva Foundation, and the Segre Research Award. AS is supported by
  grants from the BSF, GIF, and ISF, and the European Union’s Horizon 2020
  Marie Skłodowska-Curie programme.
\end{acknowledgments}

\bibliography{TheBib}

\end{document}